\RequirePackage{lineno}
\documentclass[prl,twocolumn,nofootinbib]{revtex4-1}

\usepackage{cancel}
\usepackage{graphicx}
\usepackage{epsfig}
\usepackage{amsmath,amsfonts,amssymb}
\usepackage{t1enc}

\newcommand{\ttbar}{t \bar t}
\newcommand{\afb}{A_{FB}}
\newcommand{\ac}{A_C}
\newcommand{\mttbar}{m_{t \bar t}}

\begin{document}


\title{Collider-independent $\mathbf{t \bar t}$ forward-backward asymmetries}

\author{J. A. Aguilar-Saavedra}
\affiliation{Departamento de F\'{\i}sica Te\'orica y del Cosmos, Universidad de Granada,
 E-18071 Granada, Spain  and \\
 Instituto de F\'{\i}sica de Cantabria (CSIC-UC), Santander, Spain}
\author{A. Juste}
\affiliation{Instituci\'o Catalana de Recerca i Estudis Avan\c{c}ats (ICREA) and Institut de F\'{\i}sica d'Altes Energies (IFAE), Barcelona, Spain}


\begin{abstract}
We introduce the forward-backward asymmetries $A_u$, $A_d$ corresponding to $u \bar u,d\bar d \to t \bar t$ production, respectively, at hadron colliders. These are collider- and center-of-mass-independent observables, directly related to the forward-backward and charge asymmetries measured at the Tevatron and the LHC, respectively. We discuss how to extract these asymmetries from data. Because these asymmetries are collider-independent, their measurement at these two colliders could elucidate the nature of the anomalous forward-backward asymmetry measured at the Tevatron. Our framework also shows in a model-independent fashion that a positive Tevatron asymmetry exceeding the standard model expectation is compatible with the small asymmetry measured at the LHC. 
\end{abstract}

\maketitle

{\it Introduction.} The top quark is the heaviest elementary fermion discovered and, as such, it is expected to be a good probe for physics beyond the standard model (SM). At present, thousands of top quark pairs have been produced at the Fermilab Tevatron and the CERN Large Hadron Collider (LHC), allowing for a detailed study of its properties. So far, the most interesting deviation from the SM predictions has been found in the forward-backward (FB) asymmetry in $\ttbar$ production at the Tevatron, which is
defined by the relative difference (normalized to the total number) of events with $\cos \theta > 0$ and $\cos \theta < 0$, being $\theta$ the angle between the top quark and the incoming proton in the center-of-mass (CM) frame. The measurements from the CDF and D0 Collaborations, $\afb = 0.158 \pm 0.075$~\cite{Aaltonen:2011kc}, $\afb = 0.196 \pm 0.065$~\cite{Abazov:2011rq}, and $\afb = 0.162 \pm 0.047$~\cite{CDF8.7},  are found to be consistently above the SM expectation, $A_{FB}^\text{SM} = 0.089$~\cite{Hollik:2011ps}. At high $\ttbar$ invariant mass, $\mttbar > 450$ GeV, the CDF Collaboration measures an even larger asymmetry and the deviations with respect to the SM predictions are more significant. This fact has motivated a number of new physics proposals to accommodate these observations~\cite{Djouadi:2009nb,Barcelo:2011vk,Jung:2009jz,Cheung:2009ch,Nelson:2011us,Shu:2009xf} (see~\cite{Kamenik:2011wt} for reviews). These models predict a variety of striking new signals~\cite{Cao:2011ew}, including the observation of new particles~\cite{Gresham:2011pa}. But unfortunately, Tevatron and LHC searches have not found any of these new effects beyond the SM.

The Tevatron excess can also be tested at the LHC. At this collider the initial $pp$ state is symmetric and, therefore, the FB asymmetry vanishes. Still, a charge asymmetry $\ac$ can be measured, being this quantity the relative difference between the number of events with $|y_t| > |y_{\bar t}|$ and $|y_t| < |y_{\bar t}|$, with $y_t$ ($y_{\bar t}$) the rapidity of the top (anti)quark in the laboratory frame~\cite{Diener:2009ee}. The FB asymmetry at the parton level translates into a charge asymmetry in $pp$ collisions because the valence quarks $q=u,d$ have a larger average momentum fraction than antiquarks $\bar q$, $x_q > x_{\bar q}$, leading to a boost of the $t\bar t$ system along the direction of the incoming quark, and to a larger average rapidity for top quarks than antiquarks. (Note that $t \bar t$ production from $gg$ fusion is FB symmetric.) $A_C$ provides an independent test of FB-asymmetric new physics in $\ttbar$ production but, so far, the measurements from the ATLAS and CMS Collaborations, $\ac = -0.018 \pm 0.036$~\cite{Aad:2012ug}, $\ac = -0.013 \pm 0.041$~\cite{Chatrchyan:2011hk}, and $\ac = 0.004 \pm 0.015$~\cite{CMSdiff} are consistent (and slightly below) with the SM prediction $A_C^\text{SM} = 0.0115$~\cite{Kuhn:2011ri}. This is a quite puzzling situation, because the simplest SM extensions proposed to explain the Tevatron anomalies~\cite{Djouadi:2009nb,Barcelo:2011vk,Jung:2009jz,Cheung:2009ch,Nelson:2011us,Shu:2009xf} also predict an enhancement of $\ac$ at the LHC~\cite{AguilarSaavedra:2011hz}, and so do other more complex proposals~\cite{Davoudiasl:2011tv}. The same point applies to the yet unknown SM next-to-leading order (NLO) corrections to the asymmetries, which, if increasing $A_{FB}^\text{SM}$ to reach the CDF and D0 measurements, should likely increase $A_C^\text{SM}$ too, making it deviate further from the ATLAS and CMS measurements. On the other hand, the consistency of the experimental results disfavors an explanation of this puzzle by an individual unknown systematic error.

{\it Collider-independent asymmetries.} Although $A_{FB}$ and $A_C$ both arise from some FB asymmetry in $q \bar q \to t \bar t$, they
cannot be directly compared because they are inclusive observables averaging over all subprocesses $u \bar u,d \bar d, gg \to t \bar t$, which have different relative importance at the two colliders. Moreover, at the LHC the charge asymmetry is diluted because a sizable fraction of $q \bar q \to t \bar t$ events have the incoming quark with smaller momentum fraction than the antiquark. In this regard, $A_{FB}$ and $A_C$ can be considered as different `combinations' of the `intrinsic' asymmetries $A_u$, $A_d$ in $u \bar u,d\bar d \to t \bar t$, respectively. More specifically, $\afb$ and $\ac$ can be written in terms of these asymmetries $A_u$, $A_d$ in the form
\begin{eqnarray}
\afb & = & A_u F_u + A_d F_d \,, \notag \\
\ac & = & A_u F_u D_u + A_d F_d D_d \,.
\label{ec:AuAd}
\end{eqnarray}
Here, $F_{q}$, $q=u,d$ are the fractions of $u \bar u$ and $d \bar d$ events, respectively, and $D_{q}$ are `dilution' factors, defined as the relative difference between events with
$x_q > x_{\bar q}$ and $x_q < x_{\bar q}$. 
(In the case of $\afb$ the corresponding dilutions are very close to unity.)
Equations (\ref{ec:AuAd}) hold, with {\it the same} values of $A_{q}$ for the Tevatron and the LHC, for a fixed partonic CM energy $\hat s$ and, to a good approximation, if we restrict ourselves to a suitable bin of $\mttbar = \sqrt{\hat s}$. These equations are also valid in the SM at NLO.\footnote{At NLO there is also an asymmetry from $gq \to t \bar t q$, which amounts to 5\% of the total one at the LHC~\cite{Bernreuther:2012sx}. Ignoring this contribution in Eqs.~(\ref{ec:AuAd}) gives deviations in $A_u$, $A_d$ of the same magnitude as the deviations due to the finite $\mttbar$ bins, and much smaller than the experimental precision. In any case, additional $gq$ terms can be included in the right-hand side of the second equation.}
In particular, if $A_{FB}$ and $A_C$ are calculated at fixed NLO in perturbation theory, then $F_q$ are the leading-order (LO) $q \bar q$ fractions, while if the denominators in the definition of $A_{FB}$ and $A_C$ are calculated at NLO, so must be $F_q$.

{\it Measuring $A_u$ and $A_d$.} The extraction of the individual asymmetries $A_u$ and $A_d$ from the `total' ones $\afb$, $\ac$ can be done by exploiting the dependence of the latter on the velocity ($\beta$) of the $\ttbar$ system in the laboratory frame. This is a kinematical variable involving the relative boost of the CM and laboratory frames, independent of the parton-level CM energy $\hat s$ and opening angle $\theta$, the quantities parameterising the $2 \to 2$ process $q \bar q \to t \bar t$. Hence, for fixed $\hat s$ the asymmetry in $q \bar q \to t \bar t$ is independent of $\beta$. For arbitrary $\hat s$ there is a residual dependence of the asymmetry on $\beta$, induced by the dependence of the parton density functions (PDFs) on the momentum fractions~\cite{AguilarSaavedra:2011cp}. But, working within bins of $\mttbar$, with a width of 100 GeV or smaller, this dependence is negligible for practical purposes and Eqs.~(\ref{ec:AuAd}) hold, with $\beta$-dependent functions $F_{q}$, $D_{q}$ and  $\beta$-independent constants $A_{q}$. The former functions can be computed from Monte Carlo, allowing to obtain the latter with a fit to the $\afb(\beta)$ or $\ac(\beta)$ distributions.

\begin{figure}[htb]
\epsfig{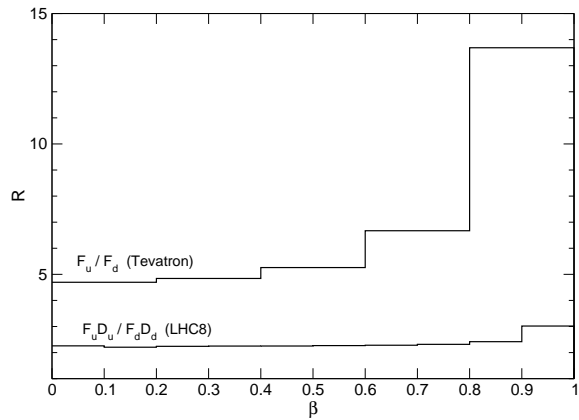}
\caption{Ratio $R$ (defined in the text) between $u$ and $d$ functions, for $400 \leq \mttbar \leq 450$ GeV.}
\label{fig:R2}
\end{figure}

The capability to discriminate $A_u$ and $A_d$ with a fit to $\afb$ ($\ac$) is driven by the variation with $\beta$ of the ratios $R=F_u /F_d$ ($R=F_u D_u /F_d D_d$). For illustration, we plot in Fig.~\ref{fig:R2} these ratios for the bin $400\leq \mttbar \leq 450$ GeV, for the Tevatron and the LHC with a CM energy of 8 TeV (LHC8) --- for 7 TeV $R$ is practically the same. Aside from the more pronounced variation of $R(\beta)$ at the Tevatron, it is worthwhile pointing out that the average value of $R$ differs by a factor $\sim 2$ at the Tevatron and the LHC. At the Tevatron both $u, d$ from the proton and $\bar u, \bar d$ from the antiproton are valence quarks, so that $d \bar d$ is roughly one quarter of the $u \bar u$ contribution. At the LHC $\bar u, \bar d$ are sea quarks and $d \bar d$ is only one half of the $u \bar u$ contribution.

In order to measure $A_{u}$ and $A_d$ from experimental data, it is necessary that the `true' fractions $F_q$ in Eqs.~(\ref{ec:AuAd}) can be well approximated by the SM ones. (The dilution factors are practically the same, at the per mille level, even with sizeable new physics contributions.)
This is a reasonable assumption since the measured $t \bar t$ differential distributions agree well with the SM prediction~\cite{Aaltonen:2009iz,CMSboost}. Second, it is also necessary that the $\mttbar$ dependence of the asymmetries, if any, is moderate, to guarantee that $A_{q}$ are in fact independent of $\beta$ within each $\mttbar$ bin. This is also fulfilled by the latest Tevatron~\cite{Abazov:2011rq,CDF8.7} and LHC~\cite{Aad:2012ug,CMSdiff} data. Under these two conditions, the values of $A_u$, $A_d$ determined from the fit effectively correspond to the FB asymmetries for $u \bar u,d \bar d \to t \bar t$, respectively. 

This setup can be explicitly tested by generating high-statistics pseudo-data samples for the Tevatron and the LHC, including a new physics contribution, and fitting $A_u$ and $A_d$ with the $F_q$, $D_q$ functions calculated from Monte Carlo. As new physics model we consider a heavy axigluon~\cite{Djouadi:2009nb}
 parametrized in the form of effective four-fermion operators with couplings $g_{11} g_{33}/\Lambda^2 = -0.93$ TeV$^{-2}$~\cite{AguilarSaavedra:2011vw}. To a good approximation, the total asymmetries are obtained by summing these new physics contributions to the SM values, which arise at one loop level. However, the inclusion of the SM contributions is not necessary for our discussion, focused on showing that $A_u$, $A_d$ are the same at Tevatron and LHC and that they can be extracted from experimental data. All computations are performed with the tree-level generator {\sc Protos}~\cite{AguilarSaavedra:2008gt}. For this benchmark model the total cross sections and new physics contributions to the asymmetries are $\sigma = 6.46$ pb, $\afb^\text{new} = 0.097$ at the Tevatron and $\sigma = 103$ (150) pb, $\ac^\text{new} = 0.02$ (0.018) at the LHC with $\sqrt{s}=7$ (8) TeV, using CTEQ6L1 PDFs~\cite{Pumplin:2002vw}.  We consider $\mttbar$ bins of 50 GeV up to 700 GeV, and for each one we perform the fits using $\beta$ bins of 0.1 (0.2) for the LHC (Tevatron). The best-fit values are presented in Table~\ref{tab:Ax}, together with the FB and charge asymmetries. 

\begin{table}[htb]
\caption{Asymmetries for the axigluon benchmark model in different $\mttbar$ bins. Rows labelled as `true' correspond to the true values, and rows labelled as `SM PDF' contain the values of $A_q$ extracted using the SM functions $F_q$, $D_q$. 
\label{tab:Ax}}
\begin{center}
\begin{tabular}{cccccccc}
\hline
\hline
& \multicolumn{3}{c}{Tevatron} & \multicolumn{3}{c}{LHC8} \\
$\mttbar$ (GeV) & $\afb$ & $A_u$ & $A_d$ & $\ac$ & $A_u$ & $A_d$ \\
\hline
$<400$ & 0.032 & 0.031 & 0.052 & 0.0051 & 0.033 & 0.042 & true\\
             &           & 0.031 & 0.052 &             & 0.033 & 0.042 & SM PDF
\\
$400-450$ & 0.068 & 0.071 & 0.083 & 0.0087 & 0.070 & 0.083 & true \\
                  &           & 0.071 & 0.083 &             & 0.069 & 0.084 & SM PDF
\\
$450-500$ & 0.106 & 0.111 & 0.122 & 0.013 & 0.113 & 0.116 & true \\
                  &           & 0.111 & 0.123 &           & 0.112 & 0.119 & SM PDF
\\
$500-550$ & 0.149 & 0.154 & 0.172 & 0.017 & 0.155 & 0.162 & true \\
                  &           & 0.154 & 0.173 &           & 0.155 & 0.168 & SM PDF
\\
$550-600$ & 0.197 & 0.201 & 0.221 & 0.022 & 0.202 & 0.212 & true \\
                  &           & 0.201 & 0.222 &           & 0.200 & 0.230 & SM PDF
\\
$600-650$ & 0.248 & 0.252 & 0.272 & 0.027 & 0.252 & 0.263 & true \\
                  &           & 0.252 & 0.274 &           & 0.249 & 0.294 & SM PDF
\\
$650-700$ & 0.301 & 0.304 & 0.334 & 0.033 & 0.305 & 0.315 & true \\
                  &           & 0.304 & 0.355 &           & 0.302 & 0.366 & SM PDF
\\
\hline
\hline
\end{tabular}
\end{center}
\end{table}

The excellent agreement between the Tevatron and LHC determinations of $A_q$ confirms that the fit indeed returns the $u\bar u$ and $d\bar d$ asymmetries at both colliders. This agreement is even more striking if we consider that, for each $\mttbar$ bin, $\afb$ and $\ac$ differ by an order of magnitude. We have also tested color octets coupling only to $u\bar u$ ($d\bar d$) as well as $Z'$ ($W'$) bosons, and checked that the fit correctly returns $A_d$ ($A_u$) consistent with zero. We also observe that using the SM fractions $F_q$ is an excellent approximation up to $\mttbar\sim 600$ GeV.  For higher $\mttbar$ correction factors could be applied using data for calibration. This sophistication is beyond the scope of the present work (the difference is smaller than the experimental sensitivity); however, we point out that the ratio between the true and SM $q \bar q$ fractions is almost independent of $\beta$ in all models tested. Thus, a global factor for each $\mttbar$ bin would suffice.

\begin{figure}[t]
\epsfig{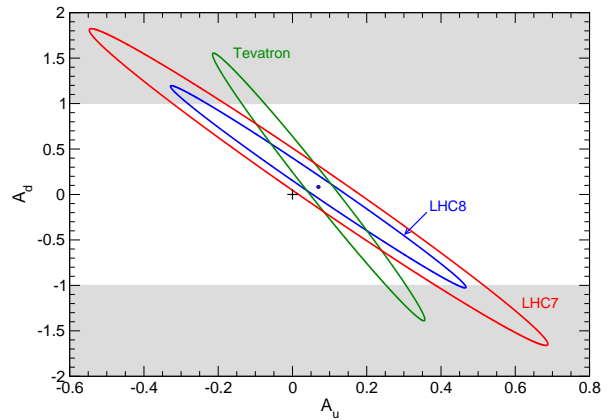}
\caption{Allowed region at 68\% C.L.  on $(A_u,A_d)$ in the $400 \leq \mttbar \leq 450$ GeV bin for the axigluon benchmark model. The dot represents the best-fit values.}
\label{fig:Amtt2}
\end{figure}

\begin{figure}[htb]
\epsfig{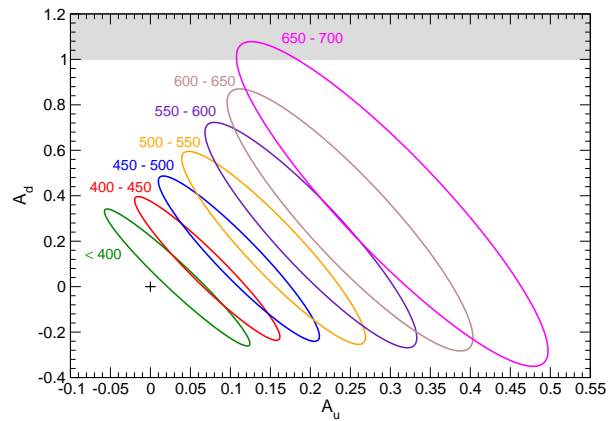}
\caption{Allowed regions at 68\% C.L. on $(A_u,A_d)$ resulting from a Tevatron-LHC combination for the axigluon benchmark model. The labels indicate the different $\mttbar$ bins (in GeV) considered.}
\label{fig:Acomb}
\end{figure}

{\it Expected sensitivity.} We now explore the prospects for the measurement of $A_u$ and $A_d$. For this purpose, an integrated luminosity of 20 fb$^{-1}$ is assumed for the Tevatron, 10 fb$^{-1}$ for the LHC with 7 TeV and 30 fb$^{-1}$ with 8 TeV, corresponding to the combination of both experiments at each collider. An overall selection efficiency of 25\% for the semileptonic $\ttbar$ decay channel is assumed, similar to that found in the experimental analyses~\cite{Aad:2012ug,Chatrchyan:2011hk}.
Apart from systematic uncertainties, which are detector-dependent and not considered here, the measurement of these asymmetries is limited by the size of the data samples. At the Tevatron, statistics at high $\beta$ are smaller because the events in $p\bar p$ collisions tend to be central. On the other hand, at the LHC the statistics are very good at high $\beta$ but the ratio $R(\beta)$ has a smaller variation than at the Tevatron, see Fig.~\ref{fig:R2}. Consequently, the limits on $(A_u,A_d)$ extracted from data are strongly anti-correlated ---these asymmetries must obey the sums in Eqs.~(\ref{ec:AuAd})--- and the resulting two-dimensional regions are very stretched ellipses, as a result of the limited statistics. For illustration, the allowed regions at 68\% confidence level (C.L.) are presented in Fig.~\ref{fig:Amtt2}, for the bin $400 \leq \mttbar \leq 450$ GeV.
The point $(0,0)$ corresponds to the SM because we are working at leading order. The shaded regions with $|A_d| > 1$ are not allowed  because these asymmetries must range between $-1$ and $1$.
Remarkably, the slopes of these ellipses differ by a factor $\sim 2$ at the Tevatron and the LHC, precisely the difference between the mean values of $R$ in Fig.~\ref{fig:R2} pointed out before. Therefore, the overlap region of the Tevatron and LHC limits is much smaller than any of them, and the combination of Tevatron and LHC measurements brings a great improvement in the determination of $A_u$ and $A_d$ (see Fig.~\ref{fig:Acomb}). 

The measurement of $A_u$ and $A_d$ may help understand the anomalous asymmetries observed at the Tevatron in two ways. A first crucial test regards the overlap between the Tevatron and LHC ellipses (that is, Fig.~\ref{fig:Amtt2} and its analogous for other $\mttbar$ bins). In our example, they intersect at the true values of $(A_u,A_d)$, as expected, by construction. But in data this is yet to be tested, and this is especially interesting having in mind the apparent tension between Tevatron and LHC measurements. A second aspect is whether the combined measurements (i.e. Fig.~\ref{fig:Acomb}) are consistent with the SM or not, what can allow to spot the presence of new physics in $\ttbar$ production. In this respect, the statistical sensitivity for the measurement of $(A_u,A_d)$  is excellent, due to the benefit from the combination of Tevatron and LHC data, as it can be seen by comparing the individual limits for $400 \leq \mttbar \leq 450$ GeV in Fig.~\ref{fig:Amtt2} and the combined one in Fig.~\ref{fig:Acomb}.

Systematic uncertainties will surely degrade the results shown here. In particular, at the LHC the charge asymmetry is small, what constitutes a difficulty for the extraction of $A_u$ and $A_d$. Still, the lower $\beta$ bins can be used for calibration because the asymmetries in these bins must be tiny, of order $10^{-4}-10^{-3}$, due to the small dilution factors and $q \bar q$ fractions at low $\beta$. Working in this direction, systematic uncertainties will likely be reduced.  Regarding the SM predictions (which must be evaluated at NLO when compared with real data), the crucial quantity to disentangle $A_u$ and $A_d$ is the ratio $R(\beta)$ because the overall normalisation of the asymmetry in a given $\mttbar$ bin is fixed by data. This ratio only depends on the PDFs for $u\bar u$ and $d\bar d$, which can be well calibrated from other processes.

\begin{figure}[htb]
\epsfig{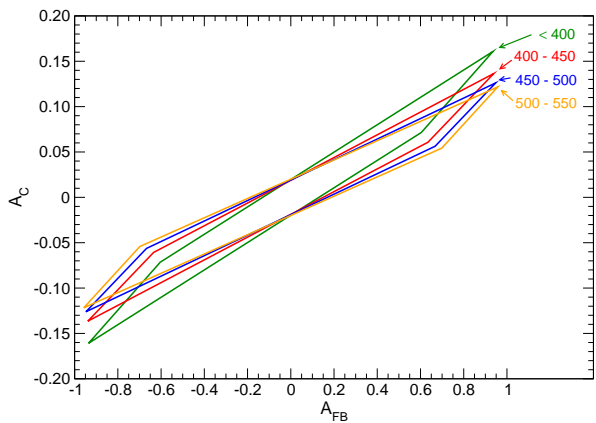}
\caption{Allowed values for $\afb$ and $\ac$, for several $\mttbar$ bins (in GeV).}
\label{fig:Apred}
\end{figure}

{\it Predictions for $A_{FB}$ and $A_C$.} As an useful by-product of our analysis, we can reverse our procedure and obtain model-independent predictions for ($\afb$,$\ac$) within each $\mttbar$ bin, by varying $A_u$ and $A_d$ between $-1$ and $1$. The resulting allowed areas are presented in Fig.~\ref{fig:Apred} for the first four $\mttbar$ bins. For higher invariant masses the allowed areas are rather similar to the one for $500-550$ GeV.
We can observe that, in each $\mttbar$ bin, it is possible to have FB asymmetries of order $0.1-0.2$ and still have a charge asymmetry close to, or even zero, at the LHC. Obviously, this is due to cancellations between $u \bar u$ and $d \bar d$ asymmetries of opposite sign, being the latter enhanced at the LHC because of the value of $R$ a factor of two smaller, see Fig.~\ref{fig:R2}. This is a notable result that shows that the Tevatron and LHC results are not incompatible but, on the contrary, their relationship deserves further investigation.

{\it Summary.} In this Letter we have introduced two collider-independent FB asymmetries $A_u$, $A_d$ in $t \bar t$ production, corresponding to the $u\bar u \to t \bar t$ and $d\bar d \to t \bar t$ subprocesses, respectively. We have discussed how they can be extracted from the measurements of the FB asymmetry in $\ttbar$ production at the Tevatron and the charge asymmetry at the LHC. We have argued how the determination of $A_u$ and $A_d$ at these two colliders can help understand the nature of the anomalous FB asymmetry observed at the Tevatron, and possibly signal new physics in $t \bar t$ production. Finally, we have used this framework to show that the asymmetry excess at the Tevatron is indeed compatible with the small charge asymmetry measured at the LHC.

\acknowledgements
We thank W. Bernreuther for useful discussions.
This work has been supported by projects FPA2006-05294 and FPA2010-17915 (MICINN),  FQM 101, FQM 03048 and FQM 6552 (Junta de Andaluc\'{\i}a).

\end{document}